\documentclass[reprint,superscriptaddress,amsmath,amssymb,aps,pra,floatfix,nofootinbib,longbibliography]{revtex4-2}
\usepackage{amsthm,booktabs}
\usepackage{microtype}
\usepackage{listings}
\usepackage[colorlinks=true,linkcolor=blue,citecolor=blue,urlcolor=blue]{hyperref}
\hypersetup{pdftitle={Local Clifford equivalence to CSS codes: Polynomial-time classification and conversion},
  pdfauthor={Mingyuan Wang and Stefan Krastanov}}
\newtheorem{theorem}{Theorem}
\newtheorem{lemma}[theorem]{Lemma}
\newtheorem{proposition}[theorem]{Proposition}
\newtheorem{corollary}[theorem]{Corollary}
\newcommand{\F}{\mathbb F_2}
\newcommand{\rank}{\operatorname{rank}}
\newcommand{\row}{\operatorname{row}}
\newcommand{\diag}{\operatorname{diag}}

\newcommand{\im}{\operatorname{im}}
\newcommand{\XX}{\mathcal X}
\newcommand{\ZZ}{\mathcal Z}
\newcommand{\HH}{\mathsf H}
\newcommand{\SSS}{\mathsf S}
\newcommand{\smallmat}[4]{\left(\begin{smallmatrix}#1&#2\\#3&#4\end{smallmatrix}\right)}

\begin{document}
\title{Local Clifford equivalence to CSS codes: Polynomial-time classification and conversion}

\author{Mingyuan Wang}
\email{mingyuanwang@umass.edu}
\affiliation{Department of Physics, University of Massachusetts Amherst}
\author{Stefan Krastanov}
\email{skrastanov@umass.edu}
\affiliation{Department of Physics, University of Massachusetts Amherst}
\affiliation{College of Information and Computer Science, University of Massachusetts Amherst}
\begin{abstract}
Calderbank--Shor--Steane (CSS) codes support decoding and fault-tolerant
logical operations through separate pure $X$ and pure $Z$ stabilizer
checks. Generator choices and single-qubit Clifford transformations can
obscure this structure. We give a deterministic polynomial-time algorithm
that decides whether an arbitrary stabilizer code is locally
Clifford-equivalent to a CSS code and constructs the required
transformation and CSS checks whenever such a form exists. For $m$
possibly redundant generator rows on $n$ qubits, the algorithm uses
$O(mn^2+n^4)$ binary field operations. We reduce the problem to affine
linear systems by characterizing local projectors that preserve the span
of the binary stabilizer labels. The conversion preserves code parameters and
check sparsity for quantum low-density parity-check (qLDPC) codes. We demonstrate CSS
recovery and decoder reuse and resolve previously open LC-to-CSS
classification questions for published code constructions. 
For graph states, LC-to-CSS equivalence corresponds to the existence of a bipartite graph in the local-complementation orbit. We derive a necessary-and-sufficient adjacency-matrix criterion for this property and construct such a graph whenever one exists.
Our algorithm makes CSS decoding and fault-tolerant
constructions accessible throughout the local Clifford equivalence
classes of CSS codes.
\end{abstract}
\maketitle

\section{Introduction}\label{sec:introduction}

Reliable quantum computation depends on error-correcting codes that
support decoding, syndrome extraction, and logical operations.
Calderbank--Shor--Steane (CSS) codes connect these tasks to classical
coding theory by separating their stabilizer checks into pure $X$ and
pure $Z$ operators~\cite{calderbankShor,steane}. CSS codes can therefore be
constructed from pairs of classical binary codes with orthogonal check
spaces, and decoded by applying classical decoders separately to the
bit- and phase-error syndromes.
The separation also permits logical
operations such as transversal CNOT between code
blocks~\cite{gottesmanFaultTolerant,albert}. These separate checks provide
the input to syndrome-extraction constructions~\cite{huangBrown}
and methods for fault-tolerant preparation of logical basis
states~\cite{pehamPreparation}.

CSS form can be hidden by generator changes and single-qubit
Clifford gates, which permute $X$, $Y$, and $Z$ up to signs while
preserving the code parameters. Such local Clifford (LC) transformations
are also used to adapt stabilizer codes to biased
noise~\cite{bonillaXZZX,duaSurface}. To use CSS decoding and fault-tolerant
methods, one must first find a suitable LC transformation and recover
pure $X$ and pure $Z$ stabilizer generators for the transformed code.
These choices are coupled: each local gate acts on every check involving
its qubit, while products of mixed generators may yield pure checks not
explicitly listed in the supplied generating set.

For a fixed LC transformation, CSS recognition reduces to a rank test on the transformed binary check matrix. Cross and Vandeth describe a
brute-force search over the $6^n$ local Clifford label assignments, applying
this rank test to determine whether a code can be transformed into CSS
form~\cite{crossVandeth}. Its exponential cost makes this search
impractical for large codes.
Recent searches for perturbed bivariate bicycle codes use LC-to-CSS
tests restricted to selected local Clifford families~\cite{cruzBenitoDiscovery}.
Polynomial-time tests for local Clifford equivalence of
two prescribed stabilizer states~\cite{vandenNestAlgorithm} take both
the source and target as input. In LC-to-CSS classification, the target CSS code is unknown. Baldelli et al. identify
general LC-to-CSS classification as an open algorithmic
problem~\cite{baldelli}.

We resolve this open problem.
We give a deterministic polynomial-time algorithm that decides whether
an arbitrary stabilizer code is locally Clifford-equivalent to a
CSS code. We allow redundant generators and any number of logical
qubits. 
Whenever such a representation exists, we return the required single-qubit Clifford transformations together with an independent generating set of pure $X$ and pure $Z$ stabilizer checks for the transformed code.
For $m$ stabilizer generators on $n$ qubits, we prove
bounds of $O(mn^2+n^4)$ binary field operations and $O(mn+n^3)$ bits of
storage; for independent generators, the time bound is $O(n^4)$. 
We obtain these guarantees by expressing CSS equivalence in terms of local rank-one projectors that preserve the linear span of the binary Pauli labels of the stabilizer generators.
We reduce the quadratic projector conditions to affine linear systems.
This reduction replaces exhaustive enumeration of local Clifford
transformations with binary linear algebra.

The recovered CSS representation makes CSS decoding and logical
operations available to the original code while preserving its
code parameters. For quantum low-density parity-check (qLDPC)
inputs~\cite{breuckmannEberhardt}, we construct CSS checks with
bounded weight and a bounded number of checks per qubit.
For the XYZ-2D family of Baldelli et al.~\cite{baldelli}, we obtain
an explicit CSS conversion for every odd distance $d\geq3$ that
preserves every check support and hence the original geometric
locality. Validation of CSS recovery and decoder reuse on
Clifford-deformed bivariate bicycle codes~\cite{bravyiBB} is
reported in Appendix~\ref{app:bb-validation}.

We also show that the 512-qubit XYZ-IS instance of Baldelli
et al.~\cite{baldelli} is not LC-equivalent to CSS, and exclude all
12 published XYZ cyclic instances~\cite{liangXYZCyclic} from the
LC orbits of CSS codes.
We resolve the 357 previously undecided entries in the perturbed
bivariate bicycle catalog~\cite{cruzBenitoDiscovery}: none is
LC-equivalent to CSS. Among the 191 symplectically halved pair-partition
entries~\cite{leeHalved,halvedCatalog}, 190 likewise lie outside the LC
orbits of CSS codes despite being constructed from CSS parents.

We also apply our algorithm to graph states, where the LC-to-CSS problem
has a graph-theoretic formulation. Local Clifford equivalence to CSS is
equivalent to the existence of a bipartite graph in the
local-complementation orbit~\cite{chenLo,vandenNest}. We derive an adjacency-matrix criterion for this property and return
a bipartite representative together with the local Clifford map to
its graph state. The bipartite representative and local map also yield
a protocol for verifying preparation of the input graph state using
at most two settings of single-qubit Pauli
measurements~\cite{dangniamVerification}.

Section~\ref{sec:setting} fixes the stabilizer conventions and states the
main result. Section~\ref{sec:structure} establishes the reduction of
LC-to-CSS equivalence to affine linear systems, which enables
polynomial-time classification and conversion.
Section~\ref{sec:algorithm} presents the algorithm, proves its correctness
and complexity bounds, and explains how its output supports decoding and
logical operations. Section~\ref{sec:published-codes} classifies published
XYZ, perturbed bivariate bicycle, and symplectically halved pair-partition
codes. Section~\ref{sec:graphs} develops the
graph-state criterion and constructs bipartite representatives together
with the corresponding local Clifford maps.
\section{Stabilizer setting and main result}\label{sec:setting}

To make the LC-to-CSS question precise, we represent stabilizer generators
by binary Pauli labels and local Clifford transformations by matrices
acting on those labels. We then state the main theorem, specifying the
conversion output and its polynomial complexity bounds.

All linear algebra is over $\F$. We use row vectors with interleaved
coordinates $(x_1,z_1,\ldots,x_n,z_n)$, where $(1,0)$, $(0,1)$, and $(1,1)$
label $X$, $Z$, and $Y$, respectively. The rows of the input matrix $B$
are the binary Pauli labels of generators of a valid stabilizer
group~\cite{gottesmanThesis}. We denote their linear span over $\F$ by $S$:
\begin{equation}\label{eq:input}
 \begin{gathered}
 B\in\F^{m\times 2n},\qquad S=\row(B),\\
 r=\dim S=n-k\leq n
 \end{gathered}
\end{equation}
In particular,
\begin{equation}\label{eq:isotropy}
 B\Omega B^{\mathsf T}=0,\qquad
 \Omega=\bigoplus_{i=1}^n\smallmat{0}{1}{1}{0}.
\end{equation}
The group is assumed to contain no $-I$. Its signs affect the choice of
stabilized eigenspace but not the binary recognition problem; their
tracking and correction are described in Appendix~\ref{app:phases}.

Let $\XX$ be the subspace in which every $z_i$ vanishes and $\ZZ$ the
subspace in which every $x_i$ vanishes. We define
\begin{equation}\label{eq:css}
 S\text{ is CSS}\quad\Longleftrightarrow\quad
 S=(S\cap\XX)\oplus(S\cap\ZZ).
\end{equation}
For any basis $G$ of $S$, write $G_X$ and $G_Z$ for the matrices obtained
by retaining its $x$ and $z$ columns. The fixed-coordinate test~\cite{crossVandeth} is
\begin{equation}\label{eq:fixed-rank}
 S\text{ is CSS}\quad\Longleftrightarrow\quad
 \rank G_X+\rank G_Z=r.
\end{equation}
Indeed, $S$ lies in the direct sum of its two projected spaces, whose
dimension is the rank sum; equality of dimensions is exactly
\eqref{eq:css}. The test allows generator changes, even when the supplied
checks are not individually pure.

A local Clifford represented by the binary matrix $Q$ acts on labels as
\begin{equation}\label{eq:local-action}
 \begin{gathered}
 S\longmapsto SQ,\qquad Q=\bigoplus_{i=1}^n Q_i,\\
 Q_i\in\mathrm{GL}(2,2)=\mathrm{Sp}(2,2).
 \end{gathered}
\end{equation}
Each of the six invertible binary $2\times2$ matrices is induced by a
single-qubit Clifford~\cite{dehaeneDeMoor}. Throughout, LC denotes local Clifford;
the graph operation of local complementation will be introduced separately
in Section~\ref{sec:graphs}. A qubit permutation preserves CSS form, so
allowing permutations does not change whether the code is LC-equivalent
to CSS.

\begin{theorem}[LC-to-CSS classification and conversion]\label{thm:main}
For a valid stabilizer input \eqref{eq:input} on $n\geq1$ qubits,
one can decide deterministically whether $SQ$ is CSS for some local map
\eqref{eq:local-action} using $O(mn^2+n^4)$ binary field operations and
$O(mn+n^3)$ bits of storage. On acceptance, the algorithm constructs
$Q_1,\ldots,Q_n$ and full-row-rank matrices $H_X,H_Z$ with $n$ columns
such that their pure $X$ and pure $Z$ label rows span exactly $SQ$, and
\begin{equation}\label{eq:output}
 H_XH_Z^{\mathsf T}=0,\qquad
 \rank H_X+\rank H_Z=r.
\end{equation}
For independent input generators, the time bound is $O(n^4)$.
\end{theorem}

Because the algorithm acts on the row space of the input, it allows
redundant and zero rows, arbitrary generator changes, and all ranks $r<n$. The two classical check spaces
in \eqref{eq:output} are orthogonal; they become full orthogonal complements
only for a stabilizer state, when $r=n$. A zero-qubit input is accepted
directly with empty output.

\section{Reduction to affine linear systems}\label{sec:structure}

Using the binary formulation above, we characterize LC-to-CSS equivalence
by the existence of a projector that preserves $S$ and has rank-one local
blocks. We then reduce the quadratic projector conditions to affine linear
systems, enabling the polynomial-time classification and conversion
algorithm developed in the next section.

\subsection{CSS form and preserving local projectors}

The linear map
\begin{equation}\label{eq:fixed-projector}
 E=\bigoplus_{i=1}^n E_i,\qquad E_i=\smallmat{1}{0}{0}{0}
\end{equation}
retains the $X$ coordinates. It is a projector on binary labels. Preservation
by $E$ is equivalent to CSS form:
\begin{equation}\label{eq:fixed-preserve}
 S\text{ is CSS}\quad\Longleftrightarrow\quad SE\subseteq S.
\end{equation}
For the reverse implication, preservation by $E$ also gives preservation
by $I+E$. Every $s\in S$ then splits as $sE+s(I+E)$, with both summands in
$S$ and with disjoint coordinate types. The decomposition of CSS codes by complementary projectors appears
in earlier algebraic descriptions~\cite{rainsNonbinary,dasuBurton}. Albert uses its local conjugates
to analyze maps between spaces already in CSS form~\cite{albert}.
To decide LC-to-CSS equivalence, we must find a suitable conjugate that
preserves $S$.

\begin{lemma}[Local projector characterization]\label{lem:projector}
The space $S$ is LC-equivalent to CSS if and only if there exists
$P=\bigoplus_i P_i$ satisfying
\begin{equation}\label{eq:projector-conditions}
 \begin{gathered}
 SP\subseteq S,\qquad P_i^2=P_i,\qquad \rank P_i=1\\
 (1\leq i\leq n).
 \end{gathered}
\end{equation}
\end{lemma}
\begin{proof}
If $SQ$ is CSS, set $P=QEQ^{-1}$. Applying
\eqref{eq:fixed-preserve} to $SQ$ gives $(SQ)E\subseteq SQ$.
Multiplying on the right by $Q^{-1}$, we obtain $SP=(SQ)EQ^{-1}\subseteq S$, and conjugation by $Q$ preserves the rank
and idempotence of each block.

Conversely, for each rank-one idempotent $P_i$, choose nonzero
\emph{column} vectors $u_i\in\im P_i$ and $w_i\in\ker P_i$, where image
and kernel here refer to multiplication on columns. They are linearly
independent, and $Q_i=[u_i\ w_i]$ satisfies
\begin{equation}\label{eq:diagonalization}
 P_iQ_i=Q_iE_i,\qquad Q_i^{-1}P_iQ_i=E_i.
\end{equation}
With $Q=\bigoplus_iQ_i$, the row-action identity is
$(SQ)E=(SP)Q\subseteq SQ$. Hence $SQ$ is CSS by
\eqref{eq:fixed-preserve}. The matrices $Q_i$ are invertible binary
label maps and can therefore be implemented by single-qubit Clifford gates.
\end{proof}

The preservation condition in \eqref{eq:projector-conditions} is linear
in the entries of $P$, whereas idempotence is quadratic. The next result
shows how the quadratic conditions can be enforced by a linear constraint
on one local block.

\subsection{Two affine branches}

Consider an $R$-dimensional stabilizer label space $S$ on $N$ qubits
that cannot be written as a direct sum over disjoint nonempty sets of
qubits. The
preprocessing in Section~\ref{sec:algorithm} reduces general inputs to
this case. Let $G\in\F^{R\times2N}$ be its reduced row-echelon form
(RREF) basis and $J$ its ordered pivot set. Parameterize all trace-one
local blocks by
\begin{equation}\label{eq:local-block}
 P_i=\begin{pmatrix}a_i&b_i\\c_i&1+a_i\end{pmatrix},
 \qquad a_i,b_i,c_i\in\F.
\end{equation}
A row $w$ belongs to $\row(G)$ exactly when $w=w[:,J]G$. Consequently,
the affine preservation system is
\begin{equation}\label{eq:pivot-preservation}
 GP=(GP)[:,J]G.
\end{equation}
Its pivot-column equations are identities, leaving $R(2N-R)$ scalar
equations in $3N$ unknowns.

\begin{proposition}[Two-branch reduction]\label{prop:branches}
Choose any anchor qubit $v$. The space $S$ is LC-equivalent
to CSS if and only if at least one of the two affine systems
\begin{align}
 GP&=(GP)[:,J]G,& b_v&=0,\label{eq:branch-b}\\
 GP&=(GP)[:,J]G,& c_v&=0\label{eq:branch-c}
\end{align}
is consistent, with the blocks parameterized by \eqref{eq:local-block}.
Every solution of either system gives a preserving local rank-one
projector.
\end{proposition}
\begin{proof}
For a binary trace-one block, direct multiplication gives
\begin{equation}\label{eq:defect}
 P_i^2+P_i=b_ic_iI_2.
\end{equation}
Here $a_i^2=a_i$ and the off-diagonal entries cancel in characteristic two.
If $P$ preserves $S$, then $P^2+P=\bigoplus_i b_ic_iI_2$ does as well.
If the coefficients $b_ic_i$ were not all equal, this map and its
complement would split $S$ over two disjoint sets of qubits.
Indecomposability therefore makes $b_ic_i$ independent of $i$, and either
anchor condition forces all of them to vanish. Thus every local block is
idempotent. An idempotent has eigenvalues in $\{0,1\}$, and trace one in
dimension two gives one of each, hence rank one. The projector
characterization proves sufficiency.

Conversely, any rank-one idempotent has trace one, so every projector
satisfying Lemma~\ref{lem:projector} has form \eqref{eq:local-block}.
It satisfies preservation and $b_vc_v=0$ by \eqref{eq:defect}. Over
$\F$, at least one of $b_v,c_v$ must vanish. Hence that projector belongs
to at least one of \eqref{eq:branch-b} and \eqref{eq:branch-c}.
\end{proof}

In Abdukhalikov and Ho's additive-to-linear code test, the endomorphism must
satisfy an irreducible quadratic polynomial, and their Algorithm~1
retains a search over the resulting null space~\cite{abdukhalikovHo}.
For LC-to-CSS classification, the argument above reduces the idempotence
condition to the two affine branches.

\section{Algorithm and complexity}\label{sec:algorithm}

We now turn this reduction into an algorithm that decides LC-to-CSS
equivalence and, on acceptance, returns the local Clifford transformation
and CSS checks. We prove its correctness and complexity bounds, then
show how the recovered representation preserves sparse checks and
supports the transfer of logical operations and decoders.

\subsection{Finding the local Clifford transformation}

A solution of either affine system yields the local Clifford blocks
through the diagonalization in \eqref{eq:diagonalization}.

The algorithm includes a preprocessing step for stabilizer label spaces
that split over disjoint sets of qubits. It identifies the indecomposable
components from a row-reduced generator matrix, applies the two-branch
test independently to each component, and combines the resulting local
Clifford maps and CSS checks. Thus inputs containing several independent
code blocks or unused qubits are handled by the same algorithm. The
decomposition and its correctness are given in
Appendix~\ref{app:preprocessing}.

Algorithm~\ref{alg:main} implements the reduction. All indices are
one-based, and all arithmetic is binary. \texttt{SolveAffine} performs
Gaussian elimination, sets free variables to zero, and returns either
one solution or \texttt{INCONSISTENT}. \texttt{RowBasis} retains the nonzero
rows of an RREF. For each local projector, its image and kernel vectors
are found by testing the three nonzero columns of $\F^2$.

\begin{figure*}
\begin{lstlisting}[caption={Deciding LC-to-CSS equivalence and returning local Clifford maps and CSS checks.},label={alg:main}]
LC_TO_CSS(B, n):
    if n == 0: return YES, empty local map, empty checks
    G := RowBasis(B)
    F := graph on qubits 1,...,n
    for each row alpha of G with pivot p_alpha:
        for each j with G[alpha,j] = 1:
            add edge {qubit(p_alpha), qubit(j)} to F
    for each connected component V of F (ignore loops):
        Gv := rows with pivot qubit in V, restricted to V
        Jv := ordered pivot columns of Gv
        P_i := [[a_i,b_i],[c_i,1+a_i]] for each i in V
        L := nonpivot equations Gv*P = (Gv*P)[:,Jv]*Gv
        v := first qubit of V
        theta := SolveAffine(L together with b_v = 0)
        if theta == INCONSISTENT:
            theta := SolveAffine(L together with c_v = 0)
        if theta == INCONSISTENT: return NO, factor V
        for each i in V:
            evaluate P_i at theta
            u := the nonzero column with P_i*u = u
            w := the nonzero column with P_i*w = 0
            Q_i := [u w], stored at the original qubit i
    T := G*(direct sum of Q_i)
    H_X := RowBasis(T[:, x columns])
    H_Z := RowBasis(T[:, z columns])
    return YES, (Q_1,...,Q_n), H_X, H_Z
\end{lstlisting}
\end{figure*}

The preservation equations in this algorithm are ordinary affine scalar
equations, with explicit coefficients given in
Appendix~\ref{app:systems}. Equivalently, if $H$ is a row basis of the
ordinary dot-product orthogonal complement of $\row(G_v)$, one may use
$G_vPH^{\mathsf T}=0$. Both formulations express the same membership
condition.

To extract the checks, write $T=GQ$. Its $X$-column projection generates
the pure $X$ part of $SQ$, and its $Z$-column projection generates the
pure $Z$ part. Keeping independent rows yields the output in
\eqref{eq:output}. Both projections are needed to span the transformed
space, since the rows of $T$ can still mix their $X$ and $Z$ components.

\subsection{Correctness and complexity}

\begin{proof}[Proof of Theorem~\ref{thm:main}]
On acceptance, Proposition~\ref{prop:branches} gives a preserving
projector on each component. The preprocessing result
\eqref{eq:factorwise} combines these projectors, and
\eqref{eq:diagonalization} shows that the constructed $Q$ takes $S$ to
CSS form. Its coordinate projections span
$SQ$, so the output rank sum is $r$. Local symplecticity preserves
isotropy, giving $H_XH_Z^{\mathsf T}=0$. Conversely, any LC-to-CSS map
provides a projector by Lemma~\ref{lem:projector}. Its restriction to
each component satisfies one of the two branches, so the algorithm cannot
reject a positive input.

For the time and storage bounds, initial reduction of an $m\times2n$ matrix
uses $O(mn^2)$ field operations. Scanning the resulting basis and finding
the qubit components costs $O(n^2)$. A component of size $N$ and rank $R\leq N$
has
\begin{equation}\label{eq:system-size}
 \begin{gathered}
 d=3N\text{ variables},\\
 K=R(2N-R)+1\leq N^2+1\\
 \text{equations per branch}.
 \end{gathered}
\end{equation}
The coefficients can be constructed in $O(N^3)$ operations: a basis
of the ordinary orthogonal complement has $2N-R$ rows, and each of the
$R(2N-R)$ equations requires $O(N)$ coefficient assignments. The pivot formulation has the
same cost, as detailed in Appendix~\ref{app:systems}.

Dense elimination processes at most $d$ pivots, updating at most $K$
rows of length $O(d)$ at each pivot. Two solves therefore cost
$O(Kd^2)=O(N^4)$ operations and $O(Kd)=O(N^3)$ bits of storage. Local
diagonalization costs $O(N)$, and the two projected check bases can be
reduced in $O(N^3)$ operations. For component sizes $N_1,\ldots,N_t$,
\begin{equation}\label{eq:complexity}
 \sum_{\ell=1}^t N_\ell^4\leq
 \left(\sum_{\ell=1}^tN_\ell\right)^4=n^4.
\end{equation}
Combining these costs proves the total time bound. Processing the components
successively requires storage for the input and output checks, the local
maps, and one coefficient matrix, bounded by $O(mn+n^3)$ bits. If the
input rows are independent, $m=r\leq n$, giving $O(n^4)$ time.
\end{proof}

These bounds count scalar operations over $\F$ and require no bit packing.
Binary isotropy may be
checked on the reduced basis in $O(n^3)$ additional operations, within
the total bound.

Each accepted $Q_i$ is implemented by a single-qubit Clifford chosen
from a fixed six-entry table. Tracking Pauli signs during conjugation and row
operations produces signed pure checks; if positive checks are desired,
one linear system supplies a product of single-qubit Pauli corrections.
Appendix~\ref{app:phases} gives the table and the phase formulas. These
steps also fit within the stated bound.

\subsection{Using the recovered CSS representation}\label{sec:uses}

The conversion preserves the code length, number of logical qubits, and
distance, while making CSS decoding methods and logical operations
available to the original code. We now show how to retain sparse checks
and use the returned local Clifford transformation to transfer these
methods to the original representation.

\paragraph{Preserving sparse checks.}
Suppose the supplied matrix $B$ has $m$ rows, each supported on at most
$w$ qubits, and each qubit belongs to at most $\Delta$ supplied checks.
Once $Q$ has been found, apply it to the \emph{original} rows of $B$ and
retain the nonzero rows of $BQE$ and $BQ(I+E)$. These are pure $X$ and
pure $Z$ checks for $SQ$, with
\begin{equation}\label{eq:sparsity}
 m_{\mathrm{CSS}}\leq2m,\qquad
 w_{\mathrm{CSS}}\leq w,\qquad
 \Delta_{\mathrm{CSS}}\leq2\Delta.
\end{equation}
Indeed, CSS form implies that both projections of each transformed row
belong to $SQ$, and their sum recovers that row, so the projected rows
generate exactly $SQ$. Local Clifford maps preserve the support of each
input check, and either projection only removes qubits from that support.
Each original check contributes at most two projected checks containing
a given qubit, which proves all three bounds. If independent checks are
desired, select an independent subset of these rows and return the
selected rows themselves. This selection replaces the final
\texttt{RowBasis} steps of Algorithm~\ref{alg:main} when sparse output is
needed. The projections cost $O(mn)$ operations and subset selection
costs $O(mn^2)$, within the stated complexity. The conversion therefore
preserves bounded check weight and a bounded number of checks per qubit
for qLDPC codes.

\paragraph{Transferring logical operations.}
Let $V=\bigotimes_iV_i$ implement the returned local map, including the
sign correction of Appendix~\ref{app:phases}, so that
$V\mathcal C=\mathcal C_{\mathrm{CSS}}$ for the input code space
$\mathcal C$. A logical operation $U$ on the recovered CSS code gives
$V^\dagger U V$ on the original code. In particular, the transversal
CNOT between two identically encoded CSS blocks~\cite{gottesmanFaultTolerant,albert} yields
\begin{equation}\label{eq:transferred-cnot}
 \begin{aligned}
 U_{\mathcal C}&=(V^\dagger\otimes V^\dagger)\,
 U_{\mathrm{CNOT}}\,(V\otimes V),\\
 U_{\mathrm{CNOT}}&=\bigotimes_{i=1}^n\mathrm{CNOT}_{i,i'}.
 \end{aligned}
\end{equation}
Each factor still acts only on matching qubits $i,i'$ in the two blocks.
Thus the transferred operation remains transversal: a fault on one pair
cannot spread to another qubit within either block. The known local basis changes similarly translate CSS encoding,
fault-tolerant preparation of logical basis
states~\cite{pehamPreparation}, and syndrome-extraction
circuits~\cite{huangBrown} to the original representation.

\paragraph{Reusing decoders.}
For Pauli noise and exact syndrome data, decoding can be transferred
entirely in classical processing. Map error labels and their
probabilities by $V$, express the syndrome in the returned CSS generator
basis, and decode using the transformed noise model. A returned Pauli
recovery $R$ becomes $V^\dagger R V$ in the original coordinates; the
stored quantum state need not be transformed. This use of CSS decoders
with transformed noise probabilities is realized for Clifford-deformed
compass codes~\cite{camposBrown}. Here our algorithm supplies
both the CSS representation and the local transformation from a general
stabilizer input.

\section{CSS recovery and classification of quantum codes}
\label{sec:published-codes}

We now apply our recognition and conversion results to published
quantum codes, addressing LC-to-CSS questions left unresolved by
restricted searches or the cost of exhaustive enumeration.
We establish LC-to-CSS inequivalence for the published XYZ-IS
instance and give an explicit conversion of the odd-distance
XYZ-2D family that preserves every check support. We also classify the
published XYZ cyclic examples and complete the LC-to-CSS classifications of the
perturbed bivariate bicycle and symplectically halved
pair-partition code catalogs.

\subsection{XYZ codes}
\label{sec:xyz}

Baldelli et al.~\cite{baldelli} distinguish between an XYZ code that
is non-CSS in its given coordinates and one that remains non-CSS
under local Pauli relabeling. Their Definition~12 calls a code
\emph{genuine under local Pauli relabeling} if no product of
single-qubit Clifford unitaries transforms it into a CSS code.
Our algorithm decides this property directly.

\subsubsection{Intersecting-subset code}
\label{sec:xyz-is}

We reconstruct the $[[512,9,d]]$ intersecting-subset (XYZ-IS) code
specified by the three subset tuples in Eq.~(102) of
Ref.~\cite{baldelli}. The original work establishes genuineness
under uniform relabeling and leaves the arbitrary local-relabeling
test unperformed because of its enumeration cost.

Both affine branches of our test are inconsistent. The published
instance is therefore genuine under local Pauli relabeling:
for every choice of single-qubit Clifford unitaries
$U_1,\ldots,U_{512}$,
\begin{equation}\label{eq:xyz-genuine}
 \left(\bigotimes_{i=1}^{512}U_i\right)
 \mathcal C_{\mathrm{XYZ-IS}}
 \quad\text{is not a CSS code}.
\end{equation}
This resolves the classification of the instance without enumerating
its $6^{512}$ Pauli-label assignments.

\subsubsection{Two-dimensional codes}
\label{sec:xyz-2d}

The XYZ-2D family of Ref.~\cite{baldelli} is obtained by local Pauli
relabeling from the XYZ$^2$ hexagonal code of Srivastava
et al.~\cite{srivastavaXYZ2}. For this family, we find a local Clifford
conversion to CSS. The algorithm returns CSS witnesses at all
tested odd distances $3\leq d\leq21$. These witnesses lead to the
following construction for the entire odd-distance family.

\begin{proposition}\label{prop:xyz-2d-css}
For every odd $d\geq3$, the standard $[[2d^2,1,d]]$ XYZ-2D code
is LC-equivalent to a CSS code admitting $d^2$ $X$-type generators
and $d^2-1$ $Z$-type generators. Their individual physical supports
coincide with those of the defining XYZ-2D checks. In particular,
the conversion preserves check weights, geometric locality, and
the number of checks incident on each qubit.
\end{proposition}

We describe the conversion in the $ZZ$-link representation of the
XYZ$^2$ code. Label the two qubits of link $(i,j)$ by $u_{i,j}$ and
$v_{i,j}$, where $0\leq i,j<d$. Its link and bulk plaquette generators
are
\begin{align}
L_{i,j}
 &= Z_{u_{i,j}}Z_{v_{i,j}},\label{eq:xyz-2d-link}\\
P_{i,j}
 &= Y_{u_{i,j}}Z_{v_{i,j+1}}X_{u_{i+1,j+1}}
    \nonumber\\
 &\qquad{}\times
    Y_{v_{i+1,j+1}}Z_{u_{i+1,j}}X_{v_{i,j}},
    \label{eq:xyz-2d-plaquette}
\end{align}
with $0\leq i,j<d-1$ for the plaquettes. Let $\HH$ be the Hadamard
gate and $\SSS=\operatorname{diag}(1,\mathrm{i})$. Apply
\begin{equation}\label{eq:xyz-2d-clifford}
\begin{gathered}
U=\bigotimes_{i,j}
  \left(U^u_{i,j}\otimes U^v_{i,j}\right),\\
(U^u_{i,j},U^v_{i,j})=
\begin{cases}
(\HH\SSS,\HH),&i+j\ \text{even},\\
(\SSS,I),&i+j\ \text{odd}.
\end{cases}
\end{gathered}
\end{equation}
Here $\HH\SSS$ applies $\SSS$ first.

Retain every link generator and replace each plaquette generator by
\begin{equation}\label{eq:xyz-2d-generator-change}
\widetilde P_{i,j}=P_{i,j}L_{i+1,j+1}.
\end{equation}
This replacement preserves the plaquette support: at the two
overlapping sites, multiplication by $Z$ exchanges the $X$ and $Y$
labels without removing either site. It is also an invertible
generator change because the link generators are retained.

For a set of qubits $A$, write $X^A$ and $Z^A$ for the corresponding
uniform Pauli products. Direct conjugation gives
\begin{align}
U L_{i,j}U^\dagger
 &=
 \begin{cases}
 X^{\operatorname{supp}L_{i,j}},&i+j\ \text{even},\\
 Z^{\operatorname{supp}L_{i,j}},&i+j\ \text{odd},
 \end{cases}\label{eq:xyz-2d-link-image}\\
U\widetilde P_{i,j}U^\dagger
 &=
 \begin{cases}
 Z^{\operatorname{supp}P_{i,j}},&i+j\ \text{even},\\
 X^{\operatorname{supp}P_{i,j}},&i+j\ \text{odd}.
 \end{cases}\label{eq:xyz-2d-plaquette-image}
\end{align}
The top and left boundary generators are similarly multiplied by
one adjacent link generator; the right and bottom boundary
generators are retained. Their signed conjugation identities and
the resulting generator counts are given in
Appendix~\ref{app:xyz-2d-conversion}, completing the proof of
Proposition~\ref{prop:xyz-2d-css}.

If $F$ denotes the local Clifford relabeling from the $ZZ$-link
representation to XYZ-2D, the conversion for XYZ-2D is $UF^{-1}$.
The resulting CSS presentation retains the original weight-two,
weight-three, and weight-six check supports. It therefore provides
a geometrically local CSS description to which the noise,
syndrome, and recovery transformations of
Section~\ref{sec:uses} apply.

\subsubsection{Cyclic codes}
\label{sec:xyz-cyclic}

Liang et al.~\cite{liangXYZCyclic} introduce a separate family of
XYZ cyclic codes and study their distance and biased-noise properties.
For nonnegative integers
$a$ and $b$, their checks are the cyclic shifts of
\begin{equation}\label{eq:xyz-cyclic-seed}
X I^b Z I^a Y I Y I^a Z I^b X,
\end{equation}
on $n=2(a+b)+7$ qubits. We reconstruct and classify all 12 instances
listed in Table~1 of Ref.~\cite{liangXYZCyclic}, with
$17\leq n\leq271$. None is LC-equivalent to a CSS code.
An extended parameter scan is reported in
Appendix~\ref{app:xyz-cyclic}.

\subsection{Perturbed bivariate bicycle codes}
\label{sec:pbb}

Cruz-Benito et al. report 368 perturbed bivariate bicycle (PBB)
code entries with $36\leq n\leq360$~\cite{cruzBenitoDiscovery}.
Their restricted LC tests identify 11 CSS-equivalent entries but
leave the full LC-to-CSS status of the remaining 357 unresolved.

We resolve the classification of all 357 previously undecided
entries: none is LC-equivalent to a CSS code
(Table~\ref{tab:pbb-classification}). Applying
Algorithm~\ref{alg:main} to the checks reconstructed from the
public polynomial tuples~\cite{pbbCatalog}, we find both affine
branches inconsistent on an indecomposable component in every
case. The 11 known positive entries therefore exhaust the
CSS-equivalent codes in the catalog.

\begin{table}
\caption{Complete LC-to-CSS classification of the public PBB
catalog~\cite{cruzBenitoDiscovery}.
We resolve the 357 previously undecided cases, showing that the
11 known CSS-equivalent entries are the only ones in the catalog.}
\label{tab:pbb-classification}
\begin{ruledtabular}
\begin{tabular}{rrrr}
$n$ & Entries & LC-equivalent to CSS & Excluded\\
\hline
36  & 47  & 1 & 46\\
72  & 34  & 2 & 32\\
108 & 107 & 3 & 104\\
144 & 66  & 1 & 65\\
180 & 64  & 2 & 62\\
360 & 50  & 2 & 48\\
\hline
Total & 368 & 11 & 357
\end{tabular}
\end{ruledtabular}
\end{table}

\subsection{Symplectically halved pair-partition codes}
\label{sec:halved-pp}

Lee et al.~\cite{leeHalved} construct compact stabilizer codes
by symplectically halving CSS pair-partition codes. The halving
identifies pairs of physical qubits in a CSS parent: the entries
of a parent $X$ check on the two members of each pair become
the $X$ and $Z$ components of a check on one qubit. This produces
a stabilizer code on half as many physical qubits. The CSS structure
of the parent does not imply that the halved code is LC-equivalent
to CSS on its own physical qubits.

We classify all 191 entries in the published plain-fold and
reversing-fold catalogs~\cite{halvedCatalog}, spanning
$48\leq n\leq504$. Exactly one entry is LC-equivalent to CSS:
the $[[110,28,10]]$ plain-fold instance already identified as
such in the source repository. The remaining 190 entries are
excluded, comprising 77 plain folds and all 113 reversing
folds (Table~\ref{tab:halved-classification}).

\begin{table}
\caption{Our classification establishes that 190 of the 191
published halved pair-partition instances~\cite{leeHalved,halvedCatalog}
lie outside the local Clifford orbits of CSS codes, despite being
constructed from CSS parents. The sole CSS-equivalent entry is
the previously identified $[[110,28,10]]$ instance.}
\label{tab:halved-classification}
\begin{ruledtabular}
\begin{tabular}{lrrr}
Construction & Entries & LC-equivalent to CSS & Excluded\\
\hline
Plain folds     & 78  & 1 & 77\\
Reversing folds & 113 & 0 & 113\\
\hline
Total           & 191 & 1 & 190
\end{tabular}
\end{ruledtabular}
\end{table}

These exclusions include the instances reported with parameters
$[[95,22,11]]$ and $[[186,65,14]]$, which are used in the
circuit-level memory simulations of Ref.~\cite{leeHalved}.

\subsection{AI-assisted implementation and verification}\label{sec:ai-use}

We used OpenAI Codex (\texttt{gpt-6-astra}) to assist with code generation,
debugging, and checks of the computational examples. A separately
developed verifier, also created with AI assistance, uses exact arithmetic
over $\F$ to check the classifications from the original stabilizer check
matrices. For positive instances, it checks orthogonality of the returned
$X$- and $Z$-check spaces and verifies that the binary labels of the
corresponding pure generators span the transformed label space. For
negative instances, it reconstructs the affine constraints and checks
contradictions in both branches.

The code, certificates, and reproducibility notes are available in the
accompanying repository~\cite{lcCssRepository}. The authors take full
responsibility for the correctness of the results.

\section{Bipartite representatives of graph-state LC orbits}\label{sec:graphs}

We now specialize the affine reduction to graph states, turning
LC-to-CSS equivalence into a criterion expressed directly in terms of
the adjacency matrix. This criterion decides whether a
local-complementation orbit contains a bipartite graph, whose graph state
is LC-equivalent to CSS. We then show how
a solution yields both a bipartite representative and the corresponding
local Clifford transformation.

A graph state on a simple graph with adjacency matrix $A$ is stabilized
by $K_i=X_i\prod_{j:A_{ij}=1}Z_j$. For graph-state calculations, we group label
coordinates as $(x_1,\ldots,x_n,z_1,\ldots,z_n)$, retaining row-vector
right action. The binary labels of these generators span
\begin{equation}\label{eq:graph-space}
 S_A=\row[I_n\mid A],\qquad A=A^{\mathsf T},\qquad A_{ii}=0.
\end{equation}
Local complementation at a vertex toggles adjacency between every pair
of its neighbors. Two graph states are LC-equivalent precisely when
their graphs are related by a sequence of these
operations~\cite{vandenNest}.

Bipartite membership of such an orbit also has an isotropic-matroid
characterization: Brijder and Traldi prove that it is equivalent to the
existence of disjoint transversals $T_1,T_2$ of $M[IAS(A)]$ with
$r(T_1)+r(T_2)=n$~\cite{brijderTraldi}. Here $M[IAS(A)]$ is the binary matroid represented by
$IAS(A)=[I\mid A\mid I+A]$. A transversal selects one column from
each qubit's triple, and $r(T)$ is the rank over $\F$ of the columns
selected by $T$.

\subsection{Direct adjacency-matrix criterion}

Applying the projector reduction to $S_A$, write $D_a=\diag(a)$ and
similarly for $b,c$. The map with trace-one local blocks takes the form
\begin{equation}\label{eq:graph-blocks}
 \begin{gathered}
 P=\begin{pmatrix}D_a&D_b\\D_c&I+D_a\end{pmatrix},\\
 [I\mid A]P=[D_a+AD_c\mid D_b+A+AD_a].
 \end{gathered}
\end{equation}
A row belongs to $S_A$ exactly when its $Z$ block is its $X$ block
multiplied by $A$. Preservation therefore reads
\begin{equation}\label{eq:graph-matrix}
 D_b+A+AD_a+D_aA+AD_cA=0.
\end{equation}
Its diagonal gives $b=Ac$, because
$(AD_cA)_{ii}=\sum_k A_{ik}c_k$ for a simple binary graph.
Setting $u=c$, the off-diagonal entries are
\begin{equation}\label{eq:graph-affine}
 A_{ij}(a_i+a_j)+\sum_k A_{ik}A_{kj}u_k=A_{ij},
 \qquad i<j.
\end{equation}
The full matrix in \eqref{eq:graph-matrix} is symmetric, so these entries
and the diagonal account for every preservation equation.

\begin{corollary}[Adjacency criterion]\label{cor:graph}
For a connected simple graph and any vertex $v$, its local-complementation
orbit contains a bipartite graph if and only if at least one of the systems
\begin{equation}\label{eq:graph-branches}
 \{\eqref{eq:graph-affine},\ u_v=0\},\qquad
 \{\eqref{eq:graph-affine},\ (Au)_v=0\}
\end{equation}
is consistent. For a disconnected graph, apply the test independently to
every connected component. Decision and construction take $O(n^4)$ field
operations and $O(n^3)$ bits of storage.
\end{corollary}
\begin{proof}
For a map that preserves $S_A$ and has trace-one local blocks,
\eqref{eq:defect} gives
$P_i^2+P_i=d_iI_2$ with $d_i=u_i(Au)_i$. The scalar mask
$D(d)=\bigoplus_i d_iI_2$ preserves $S_A$ exactly when
$AD_d=D_dA$: the two blocks of $[I\mid A]D(d)$ are $D_d$ and $AD_d$.
Entrywise, this condition says $A_{ij}(d_i+d_j)=0$, so $d$ is constant
on each connected component. Either anchor condition in
\eqref{eq:graph-branches} sets that constant to zero. Conversely, a
rank-one projector has $u_v(Au)_v=0$ and hence satisfies one
branch. Lemma~\ref{lem:projector} proves equivalence to CSS, and
Proposition~\ref{prop:bipartite} below identifies this with bipartite
orbit membership.

For a component of size $N$, there are $2N$ variables and at most
$\binom N2+1$ equations per branch. Coefficient construction takes
$O(N^3)$ operations, and two dense solves take $O(N^4)$ operations
and $O(N^3)$ bits. Sum over components as in \eqref{eq:complexity}.
The construction below requires at most $O(n^3)$ additional operations.
\end{proof}

\subsection{Constructing a bipartite representative}

\begin{proposition}[CSS states and bipartite graphs]\label{prop:bipartite}
A graph state is LC-equivalent to a CSS stabilizer state if and only if
its local-complementation orbit contains a bipartite graph~\cite{chenLo,vandenNest}. From a CSS
conversion one can construct a bipartite adjacency matrix and a local
Clifford transformation on the original labeled qubits.
\end{proposition}
\begin{proof}
For a bipartite graph with color classes $L,R$, applying Hadamards on
$R$ turns generators centered in $L$ into pure $X$ operators and those
centered in $R$ into pure $Z$ operators. Thus its graph state is
LC-equivalent to CSS.

Conversely, suppose a graph state has been converted to a CSS space
$(C_X\mid0)\oplus(0\mid C_Z)$. It has rank $n$, so orthogonality and
dimension give $C_Z=C_X^\perp$. Set $p=\dim C_X$, and choose an ordered set of coordinates
$I=(i_1,\ldots,i_p)$ such that the corresponding $p\times p$ submatrix
of a basis matrix of $C_X$ is invertible. Such a set is called an
information set. Let $J=(j_1,\ldots,j_{n-p})$ be its ordered complement.
In the coordinate listing $(I,J)$, row reduction gives the check bases
\begin{equation}\label{eq:systematic}
 H_X=[I_p\mid M],\qquad H_Z=[M^{\mathsf T}\mid I_{n-p}].
\end{equation}
The rows of the second matrix form a basis of $C_X^\perp$.
Hadamards on $J$ turn the pure rows of these two matrices into the
graph-state basis $[I_n\mid A']$, where in this listing
\begin{equation}\label{eq:bipartite-matrix}
 A'=\begin{pmatrix}0&M\\M^{\mathsf T}&0\end{pmatrix}.
\end{equation}
On the original labels, set $A'_{i_\alpha j_\beta}=M_{\alpha\beta}$,
set the transpose entries equal, and set all other entries to zero.
The choice of information set is bookkeeping, requiring no physical
permutation.

If $Q_i$ was the input-to-CSS map, the composite row label map is
\begin{equation}\label{eq:graph-witness}
 Q_i'=\begin{cases}
 Q_i,&i\in I,\\
 Q_i\smallmat{0}{1}{1}{0},&i\in J.
 \end{cases}
\end{equation}
This sends $S_A$ to $S_{A'}$. Track signs during the
conversion and final row reduction. If the standard graph generators
have minus-sign vector $\eta$, applying $\prod_i Z_i^{\eta_i}$ corrects
them, since $Z_i$ anticommutes only with the generator centered at $i$.
The resulting physical local Clifford maps the input graph state to
the positive graph state of $A'$. The graph-state equivalence
theorem~\cite{vandenNest} places $A'$ in the same local-complementation
orbit.
\end{proof}

For an accepted adjacency system, set $c=u$, $b=Au$ and construct the
projectors \eqref{eq:local-block}. Their local diagonalizations give the
initial $Q_i$, after which \eqref{eq:systematic}--\eqref{eq:graph-witness}
complete the construction. Appendix~\ref{app:systems} gives the full
graph algorithm.

The bipartition gives a verification protocol with at most two local
Pauli measurement settings~\cite{dangniamVerification}. Measuring $X$
on $I$ and $Z$ on $J$ tests the graph generators centered in $I$;
reversing the choices tests those centered in $J$. Their unique common
$+1$ eigenstate is the target graph state. Let $W=\bigotimes_i W_i$
be the returned physical Clifford mapping the input to this state.
Replacing each measured Pauli $M_i$ by $W_i^\dagger M_iW_i$ transfers
the tests to the input state, using single-qubit Pauli measurements
with known outcome-sign corrections.

\subsection{Illustrations and an immediate obstruction}

For the triangle, whose off-diagonal adjacency entries are all one,
the three equations \eqref{eq:graph-affine} are
\begin{equation}\label{eq:triangle-equations}
 \begin{aligned}
 a_1+a_2+u_3&=1,\\
 a_1+a_3+u_2&=1,\\
 a_2+a_3+u_1&=1.
 \end{aligned}
\end{equation}
Take $a=u=(1,0,0)$. Then $b=Au=(0,1,1)$, so the branch with
$(Au)_1=0$ is feasible. The projectors and a corresponding CSS conversion
are
\begin{equation}\label{eq:triangle-witness}
 \begin{aligned}
 P_1&=\smallmat{1}{0}{1}{0},&
 P_2=P_3&=\smallmat{0}{1}{0}{1},\\
 Q_1&=\smallmat{1}{0}{1}{1},&
 Q_2=Q_3&=\smallmat{1}{1}{1}{0}.
 \end{aligned}
\end{equation}
The transformed label space is spanned by the binary labels of
$XXX, ZZI, ZIZ$.
This is the three-qubit Greenberger--Horne--Zeilinger (GHZ) stabilizer;
its LC relation to the triangle also underlies the extension of
GHZ-preserving distillation circuits to
graph states~\cite{wangGHZ}. Choosing
$I=\{1\}$ gives $M=(1\ 1)$, and Hadamards on qubits 2 and 3 produce the
star with edges $12,13$. An explicit physical map from the positive
triangle graph state to the positive star graph state, up to global
phase, is
\begin{equation}\label{eq:triangle-physical}
 U=(\HH\SSS\HH)_1\otimes(Z\SSS)_2\otimes(Z\SSS)_3,
\end{equation}
where $\HH$ is the Hadamard, $\SSS=\diag(1,i)$, and the rightmost gate
acts first. Before the final $Z_2Z_3$ correction, the images of the
triangle generators are $L_1,-L_1L_2,-L_1L_3$, with $L_i$ the standard
star generators. This verifies the signs as well as the binary map.

For the five-cycle, every edge has no common neighbor. Its five edge
equations reduce to $a_i+a_{i+1}=1$, with indices modulo five.
Adding them gives $0=1$, so both anchor branches are inconsistent.
The same argument gives the following obstruction.

\begin{corollary}\label{cor:triangle-free-edges}
Let $E_0$ be the set of edges of a simple graph that lie in no triangle.
If the spanning subgraph with edge set $E_0$ contains an odd cycle, the
local-complementation orbit contains no bipartite graph. In particular,
every triangle-free nonbipartite graph is rejected.
\end{corollary}
\begin{proof}
For $ij\in E_0$, the common-neighbor sum in \eqref{eq:graph-affine}
vanishes, leaving $a_i+a_j=1$. Summing these equations around an odd
cycle is inconsistent.
\end{proof}

\section{Conclusion}\label{sec:conclusion}

We give the first polynomial-time algorithm for LC-to-CSS
classification and conversion of arbitrary stabilizer codes. By
characterizing local projectors that preserve the span of the binary
stabilizer labels, we reduce the LC-to-CSS problem to affine linear
systems. The algorithm returns a local Clifford transformation and
pure $X$ and pure $Z$ checks whenever a CSS representation exists.
A negative answer certifies that no local Clifford transformation
can map the input code to CSS form.
For graph states, our approach translates to an adjacency-matrix
criterion that decides in polynomial time whether the underlying
graph can be transformed into a bipartite graph by local
complementations. When this is possible, we construct the bipartite
graph and the local Clifford transformation between the corresponding
graph states. We also resolve previously undecided LC-to-CSS
classifications for published code constructions.

The LC-to-CSS conversion makes CSS decoding, logical operations, and
fault-tolerant constructions available across the local Clifford
equivalence class of each CSS code. The conversion preserves code
length, number of logical qubits, and distance. For qLDPC inputs,
sparse CSS checks can be constructed while retaining bounded check
weight and a bounded number of checks per qubit. CSS decoders can be
reused with the transformed noise model, and transversal logical
operations remain transversal under the returned local basis changes.
For graph states, the recovered basis changes give access to protocols
for two-colorable resources, including multipartite entanglement
purification~\cite{aschauerPurification}.
Graph-preserving circuit optimization likewise uses local Clifford
equivalence to reuse purification constructions across an LC
orbit~\cite{wangGraphPurification}.


\begin{acknowledgments}
We acknowledge support from NSF grants 1941583, 2346089, 2402861, 2522101.
We also used OpenAI Codex to assist with drafting and revising the manuscript. All AI-assisted text was reviewed and edited by the authors.
\end{acknowledgments}

\bibliography{references}

\onecolumngrid
\appendix
\section{Explicit affine systems and pseudocode}\label{app:systems}

This appendix supplies the details used to implement the algorithm in
Section~\ref{sec:algorithm} and its graph-state specialization in
Section~\ref{sec:graphs}: the preprocessing decomposition, explicit
membership equations, and graph-conversion pseudocode.

\subsection{Preprocessing over disjoint qubit sets}\label{app:preprocessing}

A factorization over qubits is detected by the coordinate masks
\begin{equation}\label{eq:mask}
 D(d)=\bigoplus_{i=1}^n d_iI_2,\qquad d\in\F^n.
\end{equation}
Preservation by a mask means that the restriction of every stabilizer
label to the selected qubits is again a stabilizer label. Such a mask
therefore separates two direct summands. Decomposition of stabilizers
over physical supports is also treated by Cross and
Vandeth~\cite{crossVandeth}.

Let $G$ be the nonzero rows of the reduced row-echelon form (RREF) of $B$,
with ordered pivot columns $J=(p_1,\ldots,p_r)$, so $G[:,J]=I_r$.
Write $q(j)=\lceil j/2\rceil$ for the qubit of scalar column $j$.
Construct a graph $F$ on the $n$ physical qubits by adding the edge
$\{q(p_\alpha),q(j)\}$ whenever $G_{\alpha j}=1$, ignoring loops.
Let $V_1,\ldots,V_t$ be its connected components, including isolated
qubits, and let $\mathcal V(V)$ denote the coordinate subspace supported
on a qubit set $V$.

\begin{lemma}[Scalar masks and physical factors]\label{lem:mask}
One has $SD(d)\subseteq S$ if and only if $d$ is constant on each $V_\ell$.
Moreover, the partition into the $V_\ell$ depends only on $S$, and
\begin{equation}\label{eq:direct-sum}
 S=\bigoplus_{\ell=1}^t S_\ell,
 \qquad S_\ell=S\cap\mathcal V(V_\ell).
\end{equation}
Each factor is indecomposable over a proper partition of its physical
qubits.
\end{lemma}
\begin{proof}
The pivot entries of a vector in $S$ determine its coefficients in the
basis $G$. For its $\alpha$th row $g_\alpha$, membership of
$g_\alpha D(d)$ therefore requires
\begin{equation}\label{eq:masked-row}
 g_\alpha D(d)=d_{q(p_\alpha)}g_\alpha.
\end{equation}
At every position with $G_{\alpha j}=1$, this equation says
$d_{q(j)}=d_{q(p_\alpha)}$. These are exactly the edge equalities in $F$,
and hence are equivalent to constancy on each component. Conversely,
those equalities imply \eqref{eq:masked-row} for every basis row and
therefore preserve $S$.

Every basis row is supported on the component containing its pivot
qubit. Grouping the rows by component proves \eqref{eq:direct-sum}; the
sum is direct because the supports are disjoint. To see that this partition
depends only on $S$, define two qubits to be equivalent when every preserving
mask has equal values on them. The first part identifies these equivalence
classes with the components of $F$: different components can be separated
by their indicator masks. This description uses only $S$. Finally, a
splitting of one factor over a proper qubit partition would produce a
preserving mask nonconstant on that component, a contradiction.
\end{proof}

An unused qubit is an isolated factor of rank zero. All factors have this
form when $S=\{0\}$. Since local block-diagonal maps preserve coordinate
supports, a preserving projector on $S$ restricts to one on each factor.
Conversely, projectors preserving the individual factors combine by direct
sum. Lemma~\ref{lem:projector} thus gives
\begin{equation}\label{eq:factorwise}
 \begin{gathered}
 S\text{ is LC-equivalent to CSS}\\
 \Updownarrow\\
 \text{every }S_\ell\text{ is LC-equivalent to CSS}.
 \end{gathered}
\end{equation}

With component sizes $N_1,\ldots,N_t$, the algorithm takes
$O(mn^2+\sum_\ell N_\ell^4)$ operations when check reduction is performed
separately on each component. A rank-zero component has no preservation
equations and is accepted by either branch. The zero-qubit case has no
components and returns empty matrices directly.

\subsection{Membership coefficients}

Let $G\in\F^{R\times2N}$ be an RREF basis with ordered pivots
$J=(p_1,\ldots,p_R)$. For each nonpivot column $j$, define a row
$h^{(j)}\in\F^{2N}$ by
\begin{equation}\label{eq:canonical-annihilator}
 h^{(j)}_j=1,\qquad h^{(j)}_{p_\beta}=G_{\beta j},
 \qquad h^{(j)}_k=0\text{ at all other positions}.
\end{equation}
These $2N-R$ independent rows form a matrix
$H\in\F^{(2N-R)\times2N}$ satisfying $GH^{\mathsf T}=0$. Thus
$\row(H)=\row(G)^\perp$ for the \emph{ordinary dot product}, and
\begin{equation}\label{eq:annihilator}
 \row(G)P\subseteq\row(G)
 \quad\Longleftrightarrow\quad GPH^{\mathsf T}=0.
\end{equation}
For the row $h^{(j)}$, the left side of its scalar membership test is
$(GP)_{\alpha j}+\sum_\beta(GP)_{\alpha p_\beta}G_{\beta j}$.
Hence these equations coincide exactly with the nonpivot equations of
\eqref{eq:pivot-preservation}. Any basis of the ordinary orthogonal complement
gives an equivalent system by a change of equation basis.

Write the local entries of rows $\alpha$ of $G$ and $\beta$ of $H$ as
$(x_{\alpha i},z_{\alpha i})$ and $(u_{\beta i},v_{\beta i})$.
The row action is
\begin{equation}\label{eq:row-action-explicit}
 (x,z)P_i=(xa_i+zc_i,\ xb_i+z(1+a_i)).
\end{equation}
The $(\alpha,\beta)$ entry of \eqref{eq:annihilator} gives
\begin{equation}\label{eq:coefficients}
 \sum_{i=1}^N\left[
 (x_{\alpha i}u_{\beta i}+z_{\alpha i}v_{\beta i})a_i
 +x_{\alpha i}v_{\beta i}b_i
 +z_{\alpha i}u_{\beta i}c_i\right]
 =\sum_{i=1}^N z_{\alpha i}v_{\beta i}.
\end{equation}
With $\theta=(a_1,b_1,c_1,\ldots,a_N,b_N,c_N)^{\mathsf T}$, this is
$L\theta=f$ with $R(2N-R)$ rows and $3N$ columns. Constructing $H$ from
\eqref{eq:canonical-annihilator} costs $O(N^2)$; writing all coefficients
in \eqref{eq:coefficients} costs $O(N^3)$.

Append a unit row in column $3v-1$ with right-hand side zero for the
$b_v$ branch, or in column $3v$ for the $c_v$ branch. Either consistent
system supplies a projector by Proposition~\ref{prop:branches}.
For a rank-zero component, $H=I_{2N}$ and the preservation system has no
equations; all variables are free before the anchor condition is added.

\subsection{Graph decision and bipartite construction}

Algorithm~\ref{alg:graph} constructs the local Clifford map on the original
qubit indices. The input is a symmetric zero-diagonal matrix. We use
the grouped coordinates $(x_1,\ldots,x_n,z_1,\ldots,z_n)$ of
Section~\ref{sec:graphs}, so $Q_i$ acts on columns $i$ and $n+i$.
The routine \texttt{LOCAL\_CLIFFORD} forms the block
\eqref{eq:local-block} and returns its image and kernel columns as in
\eqref{eq:diagonalization}.

\begin{lstlisting}[float,caption={Finding a bipartite graph and the local Clifford map to it.},label={alg:graph}]
BIPARTITE_REPRESENTATIVE(A):
    n := number of vertices
    for each connected component V:
        Av := A[V,V]
        N := number of vertices in V
        introduce variables a_1,...,a_N,u_1,...,u_N
        L := empty equation list
        for i = 1,...,N:
            for j = i+1,...,N:
                append Av[i,j]*(a_i+a_j)
                    + sum_k Av[i,k]*Av[k,j]*u_k = Av[i,j]
        theta := SolveAffine(L together with u_1 = 0)
        if theta == INCONSISTENT:
            theta := SolveAffine(L together with
                                sum_j Av[1,j]*u_j = 0)
        if theta == INCONSISTENT: return NO, component V
        (a,u) := theta
        b := Av*u
        for each local i in V:
            Q_i := LOCAL_CLIFFORD(a_i,b_i,u_i)
            store Q_i at the original vertex index
    T := [I | A]
    for i = 1,...,n:
        T[:, (i,n+i)] := T[:, (i,n+i)] * Q_i
    (H_X, I) := nonzero RREF rows of T[:,1:n] and their pivots
    J := ordered complement of I in vertices 1,...,n
    M := H_X[:,J]
    Anew := zero n by n matrix
    for alpha = 1,...,length(I):
        for beta = 1,...,length(J):
            Anew[I[alpha],J[beta]] := M[alpha,beta]
            Anew[J[beta],I[alpha]] := M[alpha,beta]
    for i = 1,...,n:
        Qnew_i := Q_i * [[0,1],[1,0]] if i in J, else Q_i
    return YES, Anew, (Qnew_1,...,Qnew_n)
\end{lstlisting}

To obtain Clifford gates, use the table in Appendix~\ref{app:phases}
and reduce the signed transformed graph checks to $[I\mid A_{\rm new}]$. Read the signs
$\eta_i$ of those rows and append $Z_i^{\eta_i}$ on each qubit.
This completes the state conversion without changing the binary maps.
When $n=0$, the loops and outputs are empty. When $C_X$ is zero or has
dimension $n$, the systematic matrices have an empty block $M$, and
the same construction gives the edgeless graph.

\section{Clifford gates and Pauli signs}\label{app:phases}

The main algorithm determines local maps on binary Pauli labels. Here we
give the corresponding single-qubit Clifford operations and track the
stabilizer signs, so that the binary conversion specifies a map between
the physical code spaces.

\subsection{Six local binary maps}

Let $\HH$ be the Hadamard and $\SSS=\diag(1,i)$. Table~\ref{tab:lifts}
lists the gates and their binary maps under the convention
$U\sigma(x,z)U^\dagger=\pm\sigma((x,z)Q_U)$, where
$\sigma(x,z)=i^{xz}X^xZ^z$. Products of physical gates act rightmost
first. In the row convention $Q_{UV}=Q_VQ_U$.

\begin{table}
\centering
\begin{tabular}{cccc}
\toprule
$U$ & $Q_U$ & $UXU^\dagger$ & $UZU^\dagger$\\
\midrule
$I$ & $\smallmat{1}{0}{0}{1}$ & $X$ & $Z$\\
$\HH$ & $\smallmat{0}{1}{1}{0}$ & $Z$ & $X$\\
$\SSS$ & $\smallmat{1}{1}{0}{1}$ & $Y$ & $Z$\\
$\HH\SSS$ & $\smallmat{1}{1}{1}{0}$ & $-Y$ & $X$\\
$\SSS\HH$ & $\smallmat{0}{1}{1}{1}$ & $Z$ & $Y$\\
$\HH\SSS\HH$ & $\smallmat{1}{0}{1}{1}$ & $X$ & $-Y$\\
\bottomrule
\end{tabular}
\caption{Clifford gates for all six invertible local binary maps.
The displayed Pauli signs fix the conjugation convention.}
\label{tab:lifts}
\end{table}

For an $n$-qubit row $s=(x_1,z_1,\ldots,x_n,z_n)$, use the Hermitian
representative
\begin{equation}\label{eq:pauli-convention}
 \sigma(s)=i^{\nu(s)}\bigotimes_i X_i^{x_i}Z_i^{z_i},
 \qquad \nu(s)=\sum_i x_iz_i\quad\text{as an integer}.
\end{equation}
A signed generator is $(-1)^\epsilon\sigma(s)$. To conjugate by an
elementary gate at qubit $i$, use
\begin{align}
 \HH_i:&\quad (x_i,z_i)\mapsto(z_i,x_i),
 &\epsilon&\mapsto\epsilon+x_iz_i,\label{eq:phase-H}\\
 \SSS_i:&\quad (x_i,z_i)\mapsto(x_i,z_i+x_i),
 &\epsilon&\mapsto\epsilon+x_iz_i.\label{eq:phase-S}
\end{align}
The products $x_iz_i$ on the right are evaluated before updating the
label. Applying these rules gate by gate implements every Clifford in the
table.

\subsection{Generator multiplication and sign correction}

To determine the signs of the converted CSS generators, supply the signs
of the input stabilizer generators together with their binary labels $B$.
Binary recognition uses only their labels.

For binary labels $s=(x,z)$ and $t=(x',z')$, understood here through their
two coordinate projections, multiplication obeys
\begin{equation}\label{eq:pauli-product}
 \sigma(s)\sigma(t)=i^{\kappa(s,t)}\sigma(s+t),\qquad
 \kappa(s,t)=\nu(s)+\nu(t)-\nu(s+t)+2\sum_i z_ix'_i\pmod4.
\end{equation}
The label sum is binary, while the integer expression is evaluated
modulo four. For commuting labels, $\kappa$ is even. Multiplying signed
stabilizer rows therefore adds their labels and updates the sign bit by
$\epsilon+\epsilon'+\kappa/2$ modulo two. This rule follows by commuting
the $Z$ factors of the first Pauli past the $X$ factors of the second.
It supplies the signs during all row reductions and changes of basis.

Choose $r$ independent rows of $B$ to form $G_0$, with corresponding signed
generators
$g_1,\ldots,g_r$, and let $U_0$ be the tensor product of the table
entries implementing $Q$. For an independent pure-check output
$C\in\F^{r\times2n}$, row reduction determines the coefficient matrix
$K\in\F^{r\times r}$ such that
\begin{equation}\label{eq:check-coefficients}
 C=K(G_0Q).
\end{equation}
For each row $c_j$ of $C$, form the commuting product
\begin{equation}\label{eq:signed-output}
 \widetilde g_j=\prod_{\alpha=1}^r
 (U_0g_\alpha U_0^\dagger)^{K_{j\alpha}}
 =(-1)^{\epsilon_j}\sigma(c_j).
\end{equation}
The conjugation rules and \eqref{eq:pauli-product} determine
$\epsilon_j$. In particular, when $c_j$ is obtained by an $X$ or $Z$
projection, its sign is recovered from this product. For example, the
checks $+YY,+ZZ$ generate $-XX,+ZZ$, since $YY\,ZZ=-XX$.

Conjugation by a Pauli with row label $t^{\mathsf T}$ changes those signs
by $C\Omega t$. Thus positive generators are obtained by solving
\begin{equation}\label{eq:sign-correction}
 C\Omega t=\epsilon.
\end{equation}
Since $C$ has rank $r$ and $\Omega$ is invertible, $C\Omega$ has full row
rank, so every $\epsilon$ has a solution. The correction is a tensor
product of single-qubit Paulis and hence remains local Clifford.

The same coefficient calculation recovers the signs of all retained
rows of $BQE$ and $BQ(I+E)$ in Section~\ref{sec:uses}; here $K$ has one
row per retained check. Use an independent subset for
\eqref{eq:sign-correction} while retaining the sparse output rows.
Every remaining pure check is a product of independent checks of the
same type, so the correction makes these checks positive as well.
Selecting the input basis and processing at most $2m$ output rows takes
$O(mn^2+n^3)$ operations and $O(mn+n^2)$ bits, including the coefficient
matrices, signs, and final solve. For graph output the simpler
correction $Z_i^{\eta_i}$ in Section~\ref{sec:graphs} is the corresponding
special case.

\paragraph{Syndrome conversion.}
Let $s_{\mathrm{in}}$ be the column of exact syndrome bits for the chosen
signed generators $g_\alpha$, with bit one denoting eigenvalue $-1$.
For a Pauli error with row label $e$, one has
$s_{\mathrm{in}}=G_0\Omega e^{\mathsf T}$.
The corrected Clifford $V=\sigma(t^{\mathsf T})U_0$ takes the code to the
positive CSS space and maps the error label to $eQ$. Consequently,
\begin{equation}\label{eq:syndrome-conversion}
 s_{\mathrm{CSS}}=C\Omega(eQ)^{\mathsf T}
 =K G_0Q\Omega Q^{\mathsf T}e^{\mathsf T}
 =K s_{\mathrm{in}},
\end{equation}
using $Q\Omega Q^{\mathsf T}=\Omega$. This relation applies to independent
or redundant output checks. If the supplied stabilizer generators are redundant, select the syndrome
entries corresponding to $G_0$. If measured bits refer to the unsigned
Pauli representatives of these rows, first add the known input sign bits to obtain
$s_{\mathrm{in}}$. Thus the coefficient matrix used to recover output
signs also supplies the generator change required for decoder reuse.

\section{CSS conversion of the XYZ-2D family}
\label{app:xyz-2d-conversion}

We complete the construction in Section~\ref{sec:xyz-2d} by specifying
the boundary checks, their generator changes and signed Clifford
images, and the map from the $ZZ$-link representation to XYZ-2D.
Throughout, $d\geq3$ is odd and $u_{i,j},v_{i,j}$ have the labeling
used in Eqs.~\eqref{eq:xyz-2d-link}--\eqref{eq:xyz-2d-plaquette}.

\subsection{Boundary generators and their conversion}

In the $ZZ$-link representation of the XYZ$^2$ hexagonal
code~\cite{srivastavaXYZ2}, the top, right, bottom, and left boundary
generators are, respectively,
\begin{align}
T_s&=Z_{u_{0,2s+1}}Y_{v_{0,2s+2}}X_{u_{0,2s+2}},\nonumber\\
R_s&=Y_{u_{2s+1,d-1}}X_{v_{2s+1,d-1}}Z_{u_{2s+2,d-1}},\nonumber\\
B_s&=X_{v_{d-1,2s}}Y_{u_{d-1,2s}}Z_{v_{d-1,2s+1}},\nonumber\\
A_s&=Z_{v_{2s,0}}X_{u_{2s+1,0}}Y_{v_{2s+1,0}},
\label{eq:xyz-2d-boundaries}
\end{align}
where $0\leq s<(d-1)/2$. We retain every link generator and make
the replacements
\begin{equation}\label{eq:xyz-2d-boundary-change}
\widetilde T_s=T_sL_{0,2s+2},\qquad
\widetilde A_s=A_sL_{2s+1,0},
\end{equation}
along with the plaquette replacement
\eqref{eq:xyz-2d-generator-change}. The right and bottom boundary
generators are retained. Each multiplied link lies entirely within
the support of its boundary or plaquette check and meets an $X$
or $Y$ at both endpoints. Thus multiplication changes the Pauli
labels without changing the support. Since all link generators are
retained, applying the same replacements again recovers the original
generators. The generator change is therefore invertible.

For the circuit $U$ in Eq.~\eqref{eq:xyz-2d-clifford}, the relevant
single-qubit actions are
\begin{equation}\label{eq:xyz-2d-gate-actions}
\begin{array}{c|ccc}
V & VXV^\dagger & VYV^\dagger & VZV^\dagger\\
\hline
\HH\SSS & -Y & -Z & X\\
\HH  & Z  & -Y & X\\
\SSS  & Y  & -X & Z\\
I  & X  & Y  & Z
\end{array}.
\end{equation}
These actions give the bulk images
\eqref{eq:xyz-2d-link-image}--\eqref{eq:xyz-2d-plaquette-image}.
Taking the defining generators with positive signs, the boundary
images are
\begin{align}
U\widetilde T_sU^\dagger&=-Z^{\operatorname{supp}T_s},\nonumber\\
UR_sU^\dagger&=-X^{\operatorname{supp}R_s},\nonumber\\
UB_sU^\dagger&=-Z^{\operatorname{supp}B_s},\nonumber\\
U\widetilde A_sU^\dagger&=-X^{\operatorname{supp}A_s}.
\label{eq:xyz-2d-boundary-images}
\end{align}
For example, multiplying the top check by its $ZZ$ link uses
$YZ=\mathrm{i}X$ and
$XZ=-\mathrm{i}Y$, giving
\begin{equation}
\widetilde T_s
=Z_{u_{0,2s+1}}X_{v_{0,2s+2}}Y_{u_{0,2s+2}}.
\end{equation}
The first site belongs to an odd link and the last two to an even
link. Their gates send these three factors to $Z$, $Z$, and $-Z$,
respectively, yielding the first identity in
Eq.~\eqref{eq:xyz-2d-boundary-images}. The other three identities
follow from the same actions and the even parity of $d-1$.

The even and odd links contribute $(d^2+1)/2$ $X$ checks and
$(d^2-1)/2$ $Z$ checks. There are $(d-1)^2/2$ plaquettes of each
parity, and the boundaries contribute $d-1$ checks of each type.
Consequently, the total numbers are
\begin{align}
N_X&=\frac{d^2+1}{2}+\frac{(d-1)^2}{2}+(d-1)=d^2,\nonumber\\
N_Z&=\frac{d^2-1}{2}+\frac{(d-1)^2}{2}+(d-1)=d^2-1.
\label{eq:xyz-2d-check-counts}
\end{align}
The defining checks generate the $[[2d^2,1,d]]$ code and number
$2d^2-1$, so they are independent. Both the generator change and
Clifford conjugation preserve their rank. Equations
\eqref{eq:xyz-2d-link-image}, \eqref{eq:xyz-2d-plaquette-image}, and
\eqref{eq:xyz-2d-boundary-images} therefore give a CSS generating
set for the entire transformed stabilizer group. The sign correction
of Appendix~\ref{app:phases} can make these independent generators
positive by a further local Pauli, preserving every check support.

\subsection{Local relabeling to XYZ-2D}

An explicit relabeling $F$ connects this construction to the XYZ-2D
representation of Ref.~\cite{baldelli}. Its checks are uniform Pauli
checks: within each check, all nonidentity factors are $X$, all are $Y$,
or all are $Z$. Let
$C_0=X,C_1=Y,C_2=Z$, with subscripts interpreted modulo three, and
put $t=i+j$. Define the local Clifford actions, up to Pauli signs, by
\begin{equation}\label{eq:xyz-2d-homogenization}
\begin{array}{c|cc}
\text{Site} & X\ \text{maps to} & Z\ \text{maps to}\\
\hline
u_{i,j} & C_{t+1} & C_{t-1}\\
v_{i,j} & C_t     & C_{t-1}
\end{array}.
\end{equation}
The two images in each row are distinct nonidentity Paulis, so each
row specifies an invertible local binary map. On an upper site,
$Y$ maps to $C_t$; on a lower site it maps to $C_{t+1}$. Substituting
these labels into the defining generators shows that $P_{i,j}$
becomes a uniform $C_{i+j}$ check and $L_{i,j}$ becomes a uniform
$C_{i+j-1}$ check. The boundary checks $T_s,R_s,B_s,A_s$ become
uniform checks of types
\begin{equation}
C_{2s},\qquad C_{2s+d},\qquad C_{2s+d-1},\qquad C_{2s-1},
\end{equation}
respectively. Thus $F$ gives an XYZ presentation on the same
supports, and $UF^{-1}$ converts that presentation to the CSS form
above. This proves Proposition~\ref{prop:xyz-2d-css} for every odd
$d\geq3$.

The finite computations cover both the $ZZ$-link and XYZ presentations
at $d=3,5,\ldots,21$, giving 20 accepted inputs. Their saved witnesses
were checked by exact row-space equality. A separate evaluation of
the explicit construction using the $2\times2$ gate matrices verifies
all 3530 signed generator identities at these sizes, including the
boundary identities and support preservation.

\section{Finite classification of XYZ cyclic codes}
\label{app:xyz-cyclic}

Table~\ref{tab:xyz-cyclic-instances} lists the 12 parameter pairs from
Table~1 of Ref.~\cite{liangXYZCyclic} and their LC-to-CSS
classifications. Each input is generated by all cyclic shifts of
Eq.~\eqref{eq:xyz-cyclic-seed}. All 12 instances encode one logical
qubit and are excluded from the LC orbits of CSS codes.

\begin{table}
\centering
\caption{LC-to-CSS classification of the 12 published XYZ cyclic
instances~\cite{liangXYZCyclic}. Every instance remains non-CSS
under arbitrary local Clifford transformations.}
\label{tab:xyz-cyclic-instances}
\begin{ruledtabular}
\begin{tabular}{rrrrc}
$a$ & $b$ & $n$ & $k$ & LC-equivalent to CSS\\
\hline
5   & 0  & 17  & 1 & No\\
8   & 1  & 25  & 1 & No\\
13  & 2  & 37  & 1 & No\\
20  & 3  & 53  & 1 & No\\
34  & 4  & 83  & 1 & No\\
28  & 5  & 73  & 1 & No\\
54  & 6  & 127 & 1 & No\\
64  & 7  & 149 & 1 & No\\
63  & 8  & 149 & 1 & No\\
101 & 9  & 227 & 1 & No\\
122 & 10 & 271 & 1 & No\\
114 & 11 & 257 & 1 & No
\end{tabular}
\end{ruledtabular}
\end{table}

We additionally classify the grid $0\leq a,b\leq20$. Four of its
441 parameter pairs already occur in the published list, so the
union contains 449 pairs, with $7\leq n\leq271$. All 449 inputs
are LC-to-CSS negative. The grid adds 437 instances to the 12
published examples and includes codes with $k=1$ and $k=3$.

To check the input reconstruction independently, we also construct
the binary $X$ and $Z$ parts from the circulant polynomials in
Eqs.~(14)--(15) of Ref.~\cite{liangXYZCyclic}:
\begin{align}
f_X(\xi)&=1+\xi^{a+b+2}+\xi^{a+b+4}+\xi^{2(a+b)+6},\nonumber\\
f_Z(\xi)&=\xi^{b+1}+\xi^{a+b+2}+\xi^{a+b+4}
              +\xi^{2a+b+5},
\end{align}
with exponents taken modulo $n$. Substitution of the cyclic shift
matrix reproduces the Pauli-string inputs. A separate verifier
checks their commutation and dimensions, reconstructs the affine
constraints, and verifies all 898 branch contradictions together
with the required indecomposable-factor conditions.

\section{Validation on Clifford-deformed bivariate bicycle codes}
\label{app:bb-validation}

The $[[72,12,6]]$, $[[144,12,12]]$, and $[[288,12,18]]$ bivariate
bicycle (BB) qLDPC codes of Bravyi et al.~\cite{bravyiBB} serve as known CSS examples for checking the implementation. We mix their $X$-
and $Z$-type generators and apply five independently sampled local
Clifford transformations to each code. The algorithm receives only
the resulting binary checks and the qubit count; the generating
transformations are withheld.

All 15 inputs are accepted and converted to CSS form. For each input,
we check that the pure $X$ and pure $Z$ checks are orthogonal, have
combined rank $n-12$, and reproduce the transformed signed stabilizer
group after sign correction. The supplied checks have maximum weight 12, with a maximum of nine checks
per qubit; the recovered checks have weight six, with a maximum of six
checks per qubit, in every case. For the
144-qubit instance, the output consists of 72 $X$ checks and 72 $Z$
checks, retained as a redundant sparse generating set of combined rank
132. These outputs realize the sparse conversion described in
Section~\ref{sec:uses}.

We use the recovered checks with a belief-propagation decoder and
ordered-statistics postprocessing~\cite{roffeDecoder}. For independent
biased Pauli noise and exact syndromes, the returned Clifford map and
generator relations convert the error probabilities, syndrome, and
recovery between the input and CSS descriptions. Using one recovered
instance at each size, all 4518 error trials preserve both the syndrome
relations and logical success or failure, including six deliberately
inserted logical errors that remain decoding failures. The recovered
sparse checks therefore provide a working CSS decoder interface for
the deformed inputs.

\end{document}